\documentclass[epj]{svjour}
\usepackage{graphics,amssymb}
\newcommand \etc {{\it etc.} }
\newcommand \tie {{\it i.e.}}

\newcommand \kd  {\delta}
\newcommand \ra  {\rightarrow}

\newcommand \w  {\omega}

\newcommand \vk {\vec{k}}

\newcommand \vx {\vec{x}}
\newcommand \vp {\vec{p}}
\newcommand \vecr {\vec{r}}

\newcommand \ep {\epsilon}

\newcommand \p {^{\prime}}

\newcommand \x {\cdot}
\newcommand \hf {\frac{1}{2}}
\newcommand \A {\alpha}

\newcommand \lc {\langle}
\newcommand \rc {\rangle}
\newcommand \prt {\partial}
\newcommand \D {\Delta}
\newcommand \sg {\sigma}

\newcommand \nt {\noindent}

\newcommand {\llb} { \left[ \frac{\mbox{}}{\mbox{}} \right.}
\newcommand {\lrb} { \left. \frac{\mbox{}}{\mbox{}} \right] }

\newcommand \bvec{\left( \begin{array}{c} }
\newcommand \evec{\end{array} \right)}

\newcommand \bea{\begin{eqnarray} }
\newcommand \eea{\end{eqnarray} }
\newcommand \nn {\nonumber}
\newcommand {\be} {\begin{equation}}
\newcommand {\ee} {\end{equation}}

\newcommand {\mbx} {\mbox{}}

\newcommand{\ata} {&\times&}

\begin{document}
\title{Characterizations of the medium in jet quenching calculations.}
\author{A. Majumder\inst{1} 
\thanks{\emph{Present address:} Department of Physics, Ohio State University, Columbus, Ohio 43210, USA.}%
}                     
%
%
\institute{Department of Physics, Duke University, Durham, North Carolina 27708, USA.}
\date{Received: 14.09.08 / Revised version: date}
%
\abstract{
The modification of hard jets in dense matter has so far been described by four different 
formalisms based on perturbative QCD (pQCD). In these proceedings, we compare the various 
approximations made in these different schemes, especially those regarding the structure 
of the medium through which jets propagate. Following this, we highlight some of the major 
differences in the various physical processes contained in the different approaches. 
} 
\maketitle

\section{Introduction}
\label{intro}

One of the major discoveries of the heavy-ion program at the Relativistic Heavy-Ion Collider (RHIC)
has been the observed suppression of high transverse momentum (high $p_T$) hadrons when 
compared to the yield of similar hadrons in $p$-$p$ collisions (scaled up by the expected number of 
binary collisions)~\cite{Adcox:2001jp,Adler:2002xw}. In $p$-$p$ collisions, such hadrons are 
formed in the fragmentation of high $p_T$ jets produced in hard scatterings. 
The presence of a dense medium influences the space-time development of the partonic shower from
 such jets and in turn leads to a medium modification of the final fragmentation to 
hadrons~\cite{Wang:1991xy,Gyulassy:1993hr}. 

The presence of a hard jet introduces a large energy scale within the process and allows for a calculation 
of the modification using the methods of perturbative QCD (pQCD). Following the early attempts of 
Baier-Dokshitzer-Mueller-Peigne-Schiff and 
Zakharov~(BDMPS-Z)~\cite{Baier:1996kr,Baier:1998yf,Baier:1996sk,Zakharov:1996fv,Zakharov:1997uu,Zakharov:1998sv,Baier:2000mf}, such 
calculations have grown in both sophistication and in the number of different observables that they are 
applied to. The majority of current approaches to the energy loss of light partons may 
be divided into four major schemes often referred to by the names of the original 
authors:
\begin{itemize}
\item Higher Twist scheme (HT)~\cite{Guo:2000nz,Wang:2001if,Zhang:2003yn,Majumder:2004pt,Majumder:2007hx,Majumder:2007ne}
\item Path integral approach to the opacity expansion by Armesto, Salgado and Wiedemann, 
(BDMPS-Z/ASW) \cite{Wiedemann:2000ez,Wiedemann:2000za,Salgado:2002cd,Salgado:2003gb,Armesto:2004ud,Baier:1996kr,Baier:1998yf,Baier:1996sk,Zakharov:1996fv,Zakharov:1997uu,Zakharov:1998sv}
\item Finite temperature field theory approach by Arnold, Moore and Yaffe (AMY)~\cite{Arnold:2000dr,Arnold:2001ba,Arnold:2002ja,Jeon:2003gi,Turbide:2005fk}
\item Reaction Operator approach to the opacity expansion  by Gyulassy, Levai and Vitev,
(GLV)~\cite{Gyulassy:1999zd,Gyulassy:2000er,Gyulassy:2001nm,Djordjevic:2003zk,Wicks:2005gt}
\end{itemize}All these schemes utilize slightly 
different approximations regarding the various scales involved in the calculation and somewhat different 
quantitative pictures of the medium. 

It will be demonstrated in the companion publication of Ref.~\cite{bass}, that, using these different 
formalisms to compute the medium modification 
of hard jets in an identical medium leads to rather similar predictions for experimental observables. 
In these proceedings, we outline the various differences between the theoretical formulations of the 
different schemes themselves. In Sect.~\ref{basic}, we present a brief introduction to the basic formalism 
and the definition of a medium modified fragmentation function. In Sect.~\ref{single_gluon}, we present a brief 
review of how the different schemes compute the single gluon emission spectrum. In Sect.~\ref{iteration}, 
we review how a single gluon emission spectrum is iterated in the different formalisms. We present 
concluding discussions in Sect.~\ref{conclusion}.

\section{Hard scattering and the medium modified fragmentation function}
\label{basic}

In the collision of two heavy-ions, there occasionally occurs a hard scattering between two 
initial partons which leads to two back-to-back out-going partons with large transverse momentum. 
These encounter multiple scattering in the produced medium leading to a modification of the 
final distribution of hadrons emanating from these partons. 
In the computation of this modified distribution, all schemes utilize a factorized approach where the 
final cross section to produce a hadron $h$  with transverse momentum $p_T$ 
(rapidity between $y$ and $y+dy$) 
may be expressed as a convolution of initial nuclear structure functions [$G_a^A(x_a),G_b^B(x_b) $, initial state 
nuclear effects such as shadowing and Cronin effect are understood to be included] to produce 
partons with momentum fractions $x_a,x_b$, a 
hard partonic cross section to produce a high transverse momentum parton $c$ with a 
transverse momentum $\hat{p}$ and a medium 
modified fragmentation function for the final hadron [$\tilde{D}_c^h(z)$], 

\bea
\frac{d^2 \sg^h}{dy d^2 p_T} &=& \frac{1}{\pi} \int dx_a d x_b G^A_a(x_a) G^B_b(x_b) \nn \\ 
\ata \frac{d \sg_{ab \ra cX} }{d \hat{t}} \frac{\tilde{D}_c^h(z)}{z}. \label{basic_cross}
\eea

\nt
In the vicinity of mid-rapidity, $z=p_T/\hat{p}$ and $\hat{t} = (\hat{p} - x_a P)^2$  ($P$ is the 
average incoming momentum of a nucleon in nucleus A).
The entire effect of energy loss is concentrated in the  calculation of the 
modification to the fragmentation function. The four models of energy loss are in a  
sense four schemes to estimate this quantity from perturbative QCD calculations.  

While the terminology (medium modification) used to describe the change in the fragmentation 
function seems to indicate that the medium has influenced the actual process of the formation of the final hadrons 
from the partonic cloud, 
this is not the case. All computations simply describe the change in the gluon radiation spectrum 
from a hard parton due to the presence of the medium. The final hadronization of the hard parton 
is always assumed to occur in the vacuum after the parton, with degraded energy, has escaped from the medium. 
Note that some of the hard gluons radiated from the hard parton will also encounter similar ``modification'' 
in the medium and may endure vacuum hadronization after escaping from the medium. Differences 
between formalisms also arise in the inclusion of hadrons from the fragmentation of such sub-leading gluons: 
whereas in approaches which compute the change in the distribution of final partons (such as AMY) or the change in the 
distribution of final hadrons (such as HT), hadrons from sub-leading gluons are implicitly included, formalisms which compute 
the energy loss of the leading parton  (such as ASW), do not include such sub-leading corrections.

\begin{figure}
\resizebox{1.75in}{1.75in}{\includegraphics[0in,0in][5in,5in]{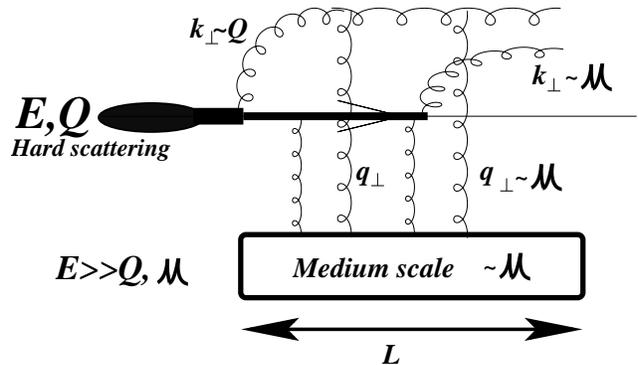}}
\caption{A schematic picture of the various scales involved in the modification of jets in dense matter. }
\label{fig0}
\end{figure}

To better appreciate the approximation schemes, one may introduce a set of scales (see Fig.~\ref{fig0}):
$E$ or $p^+$, the forward energy of the jet; $Q^2$, the virtuality of the initial jet-parton; $\mu$, 
the momentum scale of the medium and $L$, its spatial extent. Most of the   
differences between the various schemes may be reduced to the different 
relations between these various scales assumed by each scheme as well 
as by how each scheme treats or approximates the structure of the medium. In 
all schemes, the forward energy of the jet far exceeds the medium scale, $E \gg \mu$.

\section{Single gluon emission and scattering in the medium}
\label{single_gluon}

The first step in energy loss calculations is to compute the effect of a single gluon emission 
off a hard jet in the medium. The major theoretical differences between the various schemes 
arise in this calculation. It is in this step that differing assumptions regarding the medium (in different formalisms) 
are introduced. In the next section, the single emission kernel will be repeated to 
compute the effect of multiple emissions. In most cases, this involves a certain phenomenological 
picture and also introduces further differences between the different approaches. 


\subsection{Higher twist approach}
\label{HT_sec_1}


The origin of the higher twist (HT) approximation scheme lies in the calculations of 
medium enhanced higher twist corrections to the total cross section in  
Deep-Inelastic Scattering (DIS) off large nuclei~\cite{Qiu:1990xy}. In those calculations, the 
authors computed a certain class of power corrections to the total leading twist cross sections, 
which, though suppressed by powers of the hard scale $Q^2$, are enhanced by the 
extent of the medium. In the case of high $p_T$ hadron production one 
identifies and resums corrections to the single hadron inclusive cross section.

\begin{figure}
\hspace{1cm}
\resizebox{1.75in}{1.5in}{\includegraphics[0.5in,0in][4in,3.2in]{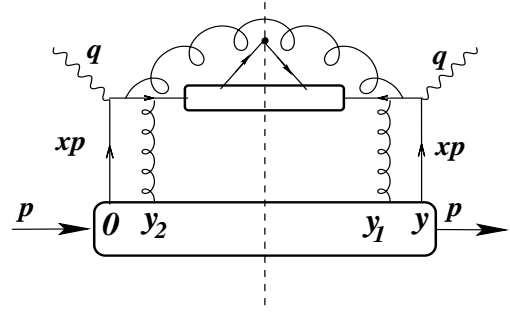}}
\caption{A typical higher twist contribution to the modification of the fragmentation 
function in medium. 
}
\label{fig1}
\end{figure}

One presupposes that the produced jet has a very large forward energy $E$ which is much 
larger than its virtuality $Q$ (which limits the transverse momentum of the radiated gluon, $k_\perp$), 
which in turn is much larger than the characteristic momentum scale 
in the medium $\mu$, i.e., $E \gg k_\perp \gg \mu$.
This hierarchy is then applied to the computation of multiple Feynman 
diagrams such as the one in Fig.~\ref{fig1}. This diagram represents 
the process of a hard virtual quark produced in a hard collision, which then radiates 
a gluon and then scatters off a soft medium gluon with transverse momentum 
$q_\perp \sim \mu$ prior to exiting the medium and fragmenting into hadrons. 
Even at the order considered, there exist various other contributions which involve scattering of 
the initial quark, off the soft gluon field, prior to radiation as well as scattering of the 
radiated gluon itself. All such contributions are combined coherently to calculate the 
modification to the fragmentation function directly.

The hierarchy of scales allows 
one to use the collinear approximation to factorize the fragmentation function and its 
modification from the hard scattering cross section. 
Thus, even though such a modified 
 fragmentation function is derived in DIS, it may be generalized to the kinematics of a 
heavy-ion collision. 
Diagrams where the outgoing parton scatters off the medium gluons, 
 such as those in Fig.~\ref{fig1},
produce a medium dependent additive contribution to the vacuum fragmentation function, 
which may be expressed as, 
\bea
\mbx && \D D_i(z,Q^2) = \int_0^{Q^2} \frac{dk_{\perp}^2}{k_{\perp}^2} 
\frac{\A_s}{2\pi} \nn \\ 
\ata\left[ \int_{z_h}^1 \frac{dx}{x} 
\sum_{j=q,g}\left\{ \D P_{i \ra j} (x,x_L,k_\perp^2) 
D_j^{h} \left(\frac{z_h}{x} \right) 
 \right\} \right].
 \label{med_mod}
\eea
\nt
In the above equation,  $\D P_{i\ra j}$ 
represents the medium modified splitting function of parton $i$ into $j$ 
where a momentum fraction $x$ is left in parton $j$. 
The new momentum fraction $x_L = k_\perp^2/(2P^-p^+ x(1-x))$ \footnote{Throughout the HT portion of these proceedings, four-vectors  
will often be referred to using the light cone convention where 
$x^+ = x^0 + x^3 $ and $x^- = ( x^0 - x^3 )/2$.}, where 
the radiated gluon or quark carries away a transverse momentum $k_\perp$, $P^-$ is 
the incoming momentum of a nucleon in the nucleus and $p$ is the momentum of the 
virtual photon. The medium modified splitting functions may be expressed as a product 
of the vacuum splitting function $P_{i \ra j}$ and a medium dependent factor, 

\bea
\D \hat{P}_{i\ra j} &=& P_{i \ra j} \int_0^L d\zeta \frac{ (N_c^2 - 1) \hat{q} }{2 \pi C_R (k_\perp^2 + 
\lc q_\perp^2 \rc) } f(\zeta,x_L) .  \label{mod_split}
\eea
\nt
Where, $C_R$ is the representation dependent Casimir and $N_c$ is the number of colours. 
The mean 
transverse momentum of the soft gluons is represented by the factor $\lc q_\perp^2 \rc$.  
The distance $\zeta$ is the distance between the origin of the jet and the location of its scattering, 
which is limited by the length of the medium $L$. The function $f(\zeta, x_L)$ depends on the 
number of scatterings per radiated gluon included and encodes the in-medium interference  
effects such as the Landau-Pomeranchuck-Migdal effect~\cite{Landau:1953um,Migdal:1956tc} .

The factor $\hat{q}$ encodes the soft gluon field in the medium, off which the jet encounters 
multiple scattering. It is given as~\cite{Majumder:2007hx}
\bea
\hat{q}(\zeta) &=& \frac{4 \pi^2 \A_s C_R}{N_c^2 - 1} 
\int \frac{d \xi^+}{2 \pi}  \frac{ d^2 \xi_\perp d^2k_\perp}{(2\pi)^2}  \label{qhat} \\
\ata \exp \left[ i \frac{q_\perp^2}{2 p^+}  \xi^+ - i \vp_\perp \x \vec{\xi}_\perp  \right] \nn \\
\ata  \lc F^{-,}_\sg (\zeta + \xi^+/2 , \vec{\xi}_\perp/2) 
F^{\sg -} (\zeta - \xi^+/2 , -\vec{\xi}_\perp/2 ) \rc.  \nn
\eea
The transport coefficient is normalized by fitting to one data point and a model such as a Woods-Saxon 
distribution for cold matter or 3-D hydrodynamical evolution for hot nuclear matter 
is invoked for its variation with space-time location. The expectation $\lc \,\, \rc$ 
is meant to be taken in the medium under consideration. Any space time dependence 
is essentially included in the implied expectation. 

The gluons which contribute to $\hat{q}$ do not have to be the entropy carriers of the system. 
In applications to cold nuclear matter, these gluons constitute the virtual gluon cloud inside the 
nucleons. In the case of a deconfined quark-gluon plasma, these may be the entropy carrying gluons 
or virtual excitations within these degrees of freedom. Which gluons the jet scatters off depends 
on the scale of the hard jet. 
It is immediately obvious from Eq.~(\ref{qhat}) that $\hat{q}$ is a function of the jet energy $p^+$. 
Note that $p^+ $ is not integrated out. The actual dependence on $p^+$ depends on the medium 
in question. In the case of confined nuclear media, or a quark gluon plasma, the dependence is 
logarithmic. 
There is also a logarithmic dependence on the virtuality 
of the jet which sets in due to radiative corrections to the definition in Eq.~(\ref{qhat}).
Also, as demonstrated in Ref.~\cite{Majumder:2006wi}, $\hat{q}$ may even possess a tensorial 
structure if the medium is not isotropic.
In the calculations of the current manuscript, both the dependence on the energy and 
virtuality of the jet will be ignored. The medium will be assumed to be isotropic. The 
values of $\hat{q}$ quoted should thus be 
considered as approximations to the full functional form.

\subsection{Opacity expansion approach}
\label{ASW_sec_1}

Unlike the higher twist scheme, which is set up to directly calculate the final 
distribution of hadrons, opacity expansion approaches such as the Gyulassy-Levai-Vitev (GLV) scheme~\cite{Gyulassy:1999zd,Gyulassy:2000er,Gyulassy:2001nm}, and the Armesto-Salgado-Wiedemann (ASW) scheme~\cite{Wiedemann:2000ez,Wiedemann:2000za,Salgado:2002cd,Salgado:2003gb,Armesto:2004ud}
were constructed primarily to deal with the 
problem of energy loss of the leading parton in dense deconfined matter. 
Both these schemes assume that the medium is composed of heavy almost static 
color scattering centers which are well separated, in the sense that the mean free path of a jet 
$\lambda \gg 1/\mu$ the color screening length of the medium~\cite{Gyulassy:1993hr}. 
The opacity of the medium $\bar{n}$, which constitutes the expansion parameter of these 
calculations, quantifies the number of scattering centers seen by a 
jet as it passes through the medium, \tie, $\bar{n} = L/\lambda$, where $L$ is the 
thickness of the medium. The difference between the two approaches of the GLV and the ASW arise 
from how these tend to expand in $n$. In the GLV formalism, one constructs a recursive operator expansion in opacity, 
whereas in the ASW approach a path integral over opacity is formulated. The solution of the recursive operator 
approach in the GLV allows for an order-by-order expansion in opacity. The path-integral in the ASW approach 
has been solved analytically in two limits: the one-scattering approximation, equivalent to a first order in opacity 
calculation in the GLV approach and in the multiple scattering approximation where all orders in opacity have been 
resummed.
In this article (as well as in the companion~\cite{bass} where results of calculation
will be compared with experimental data), the focus will lie on the path integral approach of ASW. 

The path integral approach for the energy loss of  a hard jet propagating 
in a colored medium was first introduced in Ref.~\cite{Zakharov:1996fv}.
It was later demonstrated to be equivalent to the well known BDMPS 
approach~\cite{Baier:1996kr,Baier:1998yf,Baier:1996sk} in the multiple scattering limit. 
ASW represents the current, most widespread, variant of this 
approach.
In this scheme, a hard, almost on-shell parton traversing a dense medium full of heavy scattering centers 
will engender multiple transverse scatterings of  order $\mu \ll p^+$. It will in the 
process split into an outgoing parton and a radiated gluon which will also scatter 
multiply in the medium. The radiated gluon, induced by the multiple scattering, 
has a transverse momentum $k_\perp \geq \mu$ (different from the HT approach).  
The propagation of the incoming (outgoing) partons as well as that of the radiated gluon 
in this background color field may be expressed in terms of  effective Green's functions 
[$G(\vecr_\perp,z ; \vecr_{\perp}\,\p, z\p)$ (for quark or gluon)] which obey the obvious 
Dyson-Schwinger equation, 

\bea
\mbx && G(\vecr_\perp,z ; \vec{r\p}_\perp, z\p) = G_0(\vecr_\perp,z ; \vec{r\p}_{\perp}, z\p) \nn \\
&-& i \int_z^{z\p}\!\!\!\!\!\! d \zeta \!\!\int \!\!\! d^2\vx 
G_0(\vecr_\perp, z; \vx, \zeta) A_0(\vx,\zeta) G (\vx, \zeta; \vec{r\p}_{\perp}, z\p)  \label{ASW1},
\eea
\nt
where, $G_0$ is the free Green's function and $A_0$ represents the color potential of a scattering center 
in the medium. 
The solution for the above interacting Green's function involves a path ordered Wilson line  which 
follows the potential from the location $[\vecr_{\perp}(z\p), z\p]$ to $[\vecr_{\perp}(z), z]$. Expanding 
the expression for the radiation cross section to order $A_0^{2n}$ corresponds to an expansion 
up to $n^{th}$ order in opacity.

Taking the high energy limit and the soft radiation approximation ($x\ll1$), one focuses on isolating 
the leading behavior in $x$ that arises from the large number of  interference diagrams at a given 
order of opacity. As a result of the approximations made, one recovers the BDMPS condition that 
the leading behavior in $x$ is contained solely in gluon re-scattering diagrams.  This results in the 
expression for the inclusive energy distribution for gluon radiation off an in-medium produced parton as~\cite{Wiedemann:2000za},

\bea
 x\frac{dI}{dx} &=& \frac{\A_s C_R}{(2\pi)^2 x^2} 2 {\rm Re}\!\!\! \int\limits_{\zeta_0}^\infty \!\!\!d y_l  
\!\!\int\limits_{y_l}^{\infty} \!\!\! d \bar{y}_l 
\!\!\int \!\!\! d\vec{u}\!\!\!\! \int\limits_0^{\chi x p^+} \!\!\! d \vec{k} e^{- i \vk \x \vec{u} - \hf \int d\zeta n(\zeta) \sg(\vec{u}) } \nn \\
\ata \!\frac{\prt^2}{\prt y \prt u}\!\!\!\!\!\!\!\!\int\limits_{\vec{y}=0=\vecr(y_l)}^{\vec{u}=\vecr(\bar{y})}\!\!\!\!\! \mathcal{D}r 
e^{i\int d\zeta \frac{xp^+}{2} \left( |\dot{\vecr}|^2  - \frac{n(\zeta) \sg(\vecr) }{i xp^+}\right)}, \label{ASW2}
 \eea 
\nt
where, as always, $k_\perp$ is the transverse momentum of 
the radiated gluon and $xp^+$ is its forward momentum. 
The vectors $\vec{y}$ and $\vec{u}$ represent
 the transverse locations of the emission of the gluon in 
the amplitude and the complex conjugate whereas $y_l$ and $\bar{y}_l$
 represent the longitudinal positions.  The density of 
scatterers in the medium at location $\zeta$ is $n(\zeta)$ and
 the scattering cross section is $\sg(r)$. In this form, the 
opacity is obtained as $\int n(\zeta) d \zeta$ over the extent of the medium.

Numerical implementations of this scheme have focused on
 two separate regimes. In one case, $\sg(r)$ is replaced with a
dipole form $Cr^2$ and one solves the harmonic oscillator like
 path integral. This corresponds to the case of multiple 
soft scatterings of the hard probe. 
In the limit of a static medium with a very large length, 
one obtains the simple form for the radiation distribution~\cite{Wiedemann:2000tf}, 

\bea
\w \frac{dI}{d\w} \simeq \frac{2 \A_s C_R}{\pi}  \left\{ \begin{array}{lcr} 
\sqrt{\frac{\w_c}{2\w}} \hspace{0.5cm}  &  \mbox{for} & \w < \w_c,  \\
\frac{1}{12} \left( \frac{\w}{\w_c} \right)^2 & \mbox{for} &  \w > \w_c. 
\end{array} \right. \label{omega_c}
\eea
Where $\w_c = \int d \zeta \zeta \hat{q} (\zeta)$ is called the characteristic frequency of the radiation. 
Up to constant factors, this is equal to 
mean energy lost in the medium ($\lc E \rc$) \tie, $\w_c \simeq  2 \lc E \rc / (\A_s C_R) $.
For a static medium, the integral defining $\w_c$ may be performed to obtain $\w_c = \hat{q} L^2/2$, where 
$L$ is the length of the medium and $\hat{q}$ is the jet transport coefficient, defined as the transverse momentum 
picked up by a hard jet per unit length. In actual numerical implementations, the mean $\hat{q}$ (or the $\hat{q}$ at a well 
defined location and time) is the only tunable parameter when comparing with experimental data.
For a dynamical medium of finite extent, the characteristic 
frequency and the overall mean transverse momentum gained by the jet $\lc \hat{q} L \rc $ will have to be 
estimated based on an Ansatz for the space time distribution of the transport parameter $\hat{q}$ (see Ref.~\cite{bass} for 
further details).

In the other extreme, one expands 
the exponent as a series in $n\sigma$; keeping 
only the leading order term corresponds to the picture of gluon 
radiation associated with a single scattering. In this second form, the analytical 
results of the ASW scheme formally approach those of the GLV reaction 
operator expansion~\cite{Wiedemann:2000tf}.  
In either 
case, the gluon emission intensity distribution has been found to 
be rather similar, once scaled with the characteristic 
frequency in each case.


\subsection{Finite temperature field theory approach}
\label{AMY_sec_1}


In this scheme, often referred to as the Arnold-Moore-Yaffe (AMY) approach, the energy loss of hard 
jets is considered in an 
extended medium in equilibrium at asymptotically high temperature $T \ra \infty$. 
Due to asymptotic freedom,  the coupling constant 
$g \ra 0$ at such high temperatures and a power counting scheme emerges from the 
ability to identify  a hierarchy of parametrically separated  
scales $T \gg gT \gg g^2 T$ \etc
In this limit, it becomes possible to construct an effective field theory of soft modes, \tie, $p \sim gT$ by 
summing contributions from  hard loops with $p \sim T$, into effective propagators and vertices~\cite{Braaten:1989kk}. 

\begin{figure}
\resizebox{2.in}{1.25in}{\includegraphics[0.0in,-0.5in][6in,2.5in]{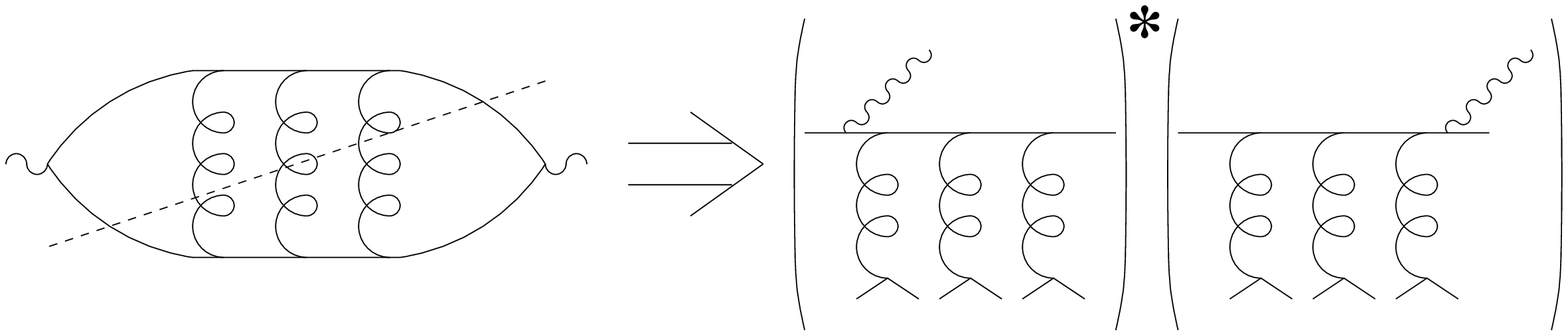}}
\caption{
A typical cut diagram in the AMY formalism.}
\label{fig2}
\end{figure}

One assumes a hard on-shell parton, with energy several 
times that of the temperature, traversing such a 
medium, undergoing soft scatterings with momentum transfers $\sim gT$ off other hard 
partons in the medium. Such soft scatterings induce collinear radiation from the parton, with 
a transverse momentum of the order of $g T$. The formation time for such collinear 
radiation $\sim 1/(g^2T) $ is of the same order of magnitude as the mean free time 
between soft scatterings~\cite{Arnold:2001ba}. As a result, multiple scatterings of the 
incoming (outgoing) parton 
and the radiated gluon need to be considered to get the leading order gluon radiation rate. 
One essentially calculates the imaginary parts of infinite order ladder diagrams such as 
those shown in Fig.~\ref{fig2}; this is done by means of integral 
equations~\cite{Arnold:2002ja}.

The imaginary parts of such ladder diagrams yield  the $1\ra2$ decay 
rates of a hard parton $(a)$ into a radiated gluon and another parton $(b)$ $\Gamma_{bg}^a$.  These 
decay rates are then used to evolve hard quark and gluon distributions from the initial hard 
collisions,  
when they are formed, to the time when they exit the medium, by means of a Fokker-Planck 
like equation~\cite{Jeon:2003gi}, which is written schematically as, 

\bea
\frac{d P_a(p)}{d t} &=& \int dk \sum_{b,c} \left[ P_b(p+k) \frac{d\Gamma^b_{ac}(p+k,p)}{dk dt} \right. \nn \\
&-&   \left. P_a(p) \frac{d \Gamma^a_{bc}(p,k)}{dk dt}  \right] . \label{AMY1}
\eea

The use of an effective theory for the description of the medium and the propagation of the jet, makes this approach
considerably more systematic than the two previous approaches: both the properties of the jet and the medium are
described using the same hierarchy of scales. It remains the only approach to date which naturally includes partonic
feedback from the medium, \tie, processes where a thermal quark or gluon may be absorbed by the hard jet
\footnote{While an attempt to include such effects in the higher twist formalism have been made in
Ref.~\cite{Wang:2001cs}, these remain as phenomenological extensions and have not been included in this manuscript.}.
In contrast to ASW and HT, this approach also (naturally) includes flavor changing interactions in the medium. 
Elastic energy loss may also be incorporated within the same basic formalism~\cite{Qin:2007rn}. However, since the 
 AMY scheme assumes a thermalized partonic medium, its applicability is somewhat limited: 
It cannot compute the quenching of jets in the confined sector. 
As a result, energy loss in cold confined nuclear matter as well as in the hadronic phase of heavy-ion collisions 
cannot be computed in this model. 
The off-shellness or 
virtuality of all jets is considered to be similar to that of hard partons in the medium, as a result, interference 
between vacuum and medium induced radiations is also not considered. 

In realistic calculations, the temperature of the medium is usually set by the underlying hydrodynamic 
simulation (see Ref.~\cite{bass} for details). While in the HT or ASW formalisms, an Ansatz is made for the 
one tunable parameter $\hat{q}$ and its relation to $T$, 
in the AMY formalism, $\hat{q}$ may be calculated directly from a knowledge of 
the temperature and the strong coupling constant $g$ (or $\A_s$). 
This is due to the precise picture of the medium used: that of a hot plasma of quarks and gluons. 
In realistic simulations, the coupling is, 
in principle,  
unknown and becomes the primary fit parameter. This is then fit by comparison to one data point.

\section{Multiple gluon emissions}
\label{iteration}

In the preceding section, the effect of a single gluon emission, stimulated by scattering in the medium, was 
considered. In order to compute the final spectrum of hadrons, this single emission kernel has to be 
be repeated to account for multiple gluon emissions and folded with a non-perturbative fragmentation function. 
Even in this procedure, the different schemes employ different methods: 
The HT scheme starts with a fragmentation 
function at a lower scale $\mu$ and evolves this distribution up to a higher scale. The ASW scheme, considers
a finite energy lost by the leading parton in multiple unrelated events by means of a Poisson distribution and folds 
the outgoing parton with a vacuum fragmentation function with a shifted momentum fraction $z$. 
The AMY formalism considers the evolution of an initial distribution of hard partons with time in the 
medium using the Fokker-Planck equation afforded by Eq.~\ref{AMY1} and final also uses a vacuum fragmentation 
function with a shifted momentum fraction $z$.

\subsection{Higher twist scheme}

In subsection~\ref{HT_sec_1}, the medium modified fragmentation function 
calculated in Eq.~(\ref{single_gluon}) included only one gluon emission 
in the medium. Any remaining gluon emissions occurred in the vacuum and 
were included in the renormalization of the vacuum fragmentation function. 
Unlike the remaining formalisms, the results from just the single gluon 
emission in the medium yield a medium modified fragmentation function 
and are already comparable with experiment. 

In reality, one expects multiple emissions to occur in the medium, followed 
by escape into the vacuum and further emissions in the vacuum. One starts 
with a vacuum fragmentation function at a low scale $\mu_{low}$ and insists
that the parton exits the medium with a certain virtuality $\mu$. Emissions from 
the scale $\mu_{low}$ up to the scale $\mu$ may be included by using the 
standard vacuum Dokshitzer-Gribov-Lipatov-Altarelli-Parisi (DGLAP) evolution 
equations~\cite{Gribov:1972ri,Dokshitzer:1977sg,Altarelli:1977zs}. 

Emissions in the medium account for the remaining evolution from the scale $\mu$ up to the 
scale $Q$.
To compute this in-medium evolution, the medium modified fragmentation function from 
single gluon emission in Eq.~(\ref{single_gluon}) is now generalized to an 
evolution equation in virtuality of the propagating parton (see Ref.~\cite{maj08} for details), i.e, 
\bea
\frac{\prt {D_q^h}(z,M^2\!\!,p^+)|_{\zeta_i}^{\zeta_f}}{\prt \log(M^2)} \!\!\!&=&\int\limits_z^1dy 
\int\limits_{\zeta_i}^{\zeta_f} d \zeta P_{i \ra j} \frac{ (N_c^2 - 1) \hat{q} (\zeta) }{2 \pi C_R (k_\perp^2 + 
\lc q_\perp^2 \rc) } \nn \\ 
\ata  f(\zeta,x_L,y) {D_q^h}\left. \left(\frac{z}{y},M^2\!\!,q^-y\right) \right|_{\zeta}^{\zeta_f}.  \label{in_medium_evol_eqn}
\eea

The initial conditions to this differential equation are provided by the fragmentation functions at the scale $\mu$. 
The final resulting medium modified fragmentation functions includes both vacuum and in-medium 
induced emissions from the scale $Q$ down to the scale $\mu$. Further emissions occur solely 
in the vacuum. This medium modified fragmentation function may now be convoluted with the  
cross section to produce a hard parton [as in Eq.~(\ref{basic_cross})] to find the final distribution of 
hadrons. Computation of the medium modified fragmentation function in an evolving medium 
such as the deconfined matter formed in a quark-gluon plasma involves further calculational 
details presented in Ref.~\cite{bass}.  As the HT formalism is setup to directly calculate the 
final modified fragmentation function, it offers the simplest and most direct extension to the study of multi-hadron 
observables~\cite{Majumder:2004wh}.


\subsection{Opacity expansion scheme}


In subsection~\ref{ASW_sec_1}, the opacity expansion was used to calculate
the differential spectrum for single gluon radiation from a hard parton. The calculations were 
carried out in the soft gluon limit \tie, $\w \ra 0$. The next step is 
to calculate the probability for the leading hard parton to radiate off a 
finite energy $\D E  = \ep P^+$. After losing this energy, the degraded hard parton escapes the medium 
and fragments in vacuum into a shower of hadrons.

For the parton to lose a finite fraction of its forward energy, multiple gluon emissions are required. 
Each such emission at a given opacity is assumed to be independent and a 
probabilistic scheme is set up, wherein, the jet loses an energy fraction $\D E$ in 
$n$ tries with a Poisson distribution~\cite{Salgado:2003gb,Eskola:2004cr}, 
\bea
 P_n(\ep,P^+) = \frac{e^{-\lc N_g \rc} }{n!} \Pi_{i=1}^n \llb \int d x_i \frac{dN_g}{dx_i} \lrb 
\kd(\ep - \sum_{i=1}^{n} x_i  ) , \nn \\
 \label{GLV3}
\eea
where, $\lc N_g \rc$ is the mean number of gluons radiated per coherent interaction set.
Summing over $n$ gives the probability  $P(\ep)$ for  an incident jet to lose 
a momentum fraction $\ep$ due to its passage through the medium. 

This probability distribution is then used to model a medium 
modified fragmentation function, by shifting the energy fraction available to 
produce a hadron as well as accounting for the phase space available after energy loss 
(The fragmentation function used is a vacuum fragmentation function).
The medium modified fragmentation function is thus defined as~\cite{Salgado:2003gb,Eskola:2004cr},
\bea
\tilde{D}(z,Q^2) = \int_0^{1} d\ep  P(\ep) \frac{D\left( \frac{z}{1-\ep},Q^2\right)}{1-\ep}.  \label{GLV4}
\eea
\nt
The above, modified fragmentation function is then used in a factorized formalism as in Eq.~(\ref{basic_cross}) 
to calculate the final hadronic spectrum. Additional details related to the computation of the energy loss 
probability distribution are given in Ref.~\cite{bass}.

\subsection{Finite temperature field theory scheme}

In subsection~\ref{AMY_sec_1}, the computation of the rates of parton splitting and merging in a thermalized 
deconfined medium were calculated. Based on these rates, a Fokker-Plank equation was motivated which 
computed the change in the distribution of hard partons with time spent propagating through a medium. 
If the initial distribution is taken from the cross-section to produce a hard parton as in Eq.~(\ref{basic_cross}) 
and the time spent in the medium is estimated based on the production point and the direction of propagation, 
this equation will yield the distribution of hard partons as they exit the deconfined medium. In the AMY scheme,  
these partons are no longer expected to interact with the hadronic plasma and thus do not lose energy in the 
hadronic phase. 

The final hadron spectrum at high $p_T$ is obtained by the fragmentation of jets in the vacuum after their passing
through the medium. In this approach, one calculates the medium modified fragmentation function by convoluting the
vacuum fragmentation functions with the hard parton distributions, at exit, to produce the final hadronic
spectrum~\cite{Turbide:2005fk},
\begin{eqnarray}
\label{AMY2_FF} \tilde{D}^h_{j}(z,\vec{r}_\bot, \phi) \!&=&\!\! \sum_{j'} \!\int\! dp_{j'} \frac{z'}{z} D^h_{j'}(z')
P(p_{j'}|p_j,\vec{r}_\bot, \phi). \ \ \ \ \ \
\end{eqnarray}
In the equation above, the sum over $j'$ is the sum over all parton species. The two momentum fractions are $z=p_h/p_j$ and
$z'=p_h/p_{j'}$, where $p_j$ and $p_{j'}$ are the momenta of the hard partons immediately after the hard scattering
and prior to exit from the medium and $p_h$ is the final hadron momentum. The quantity $P(p_{j'}|p_j,\vec{r}_\bot, \phi)$
represents the solution to Eq.~(\ref{AMY1}), which is the probability of obtaining a given parton $j'$ with momentum
$p_{j'}$ when the initial condition is a parton $j$ with momentum $p_j$. The above integral depends implicitly on the
path taken by the parton and the medium profile along that path, which in turn depends on the location of the origin
$\vec{r}_{\bot}$ of the jet and its propagation angle $\phi$ with respect to the reaction plane. Therefore, one must
convolve the above expression over all transverse positions $\vec{r}_{\bot}$ and directions $\phi$. Details of 
this procedure are presented in the companion paper of Ref.~\cite{bass}.

\section{Discussions and conclusions}
\label{conclusion}

In these proceedings, the different underlying theoretical mechanisms used in some of the prevalent 
jet energy loss calculations have been outlined. Specific attention was paid to how each jet 
resolves the medium and on the property of the medium which controls the modification of the hard 
jet. In all cases, the jet modification formalisms may be reduced to a form which depends on 
only one parameter: this is the transport coefficient $\hat{q}$ defined as the transverse momentum 
gained by a hard parton per unit length traversed in a dense medium. 
While in the ASW formalism, $\hat{q}$ is the sole tunable parameter, in the HT formalism, $\hat{q}$ depends on the gluon field strength correlation [Eq.~\ref{qhat}] and may be calculated from a knowledge of the temperature and the coupling constant $\A_s$ 
in the AMY formalism. 

In all cases, the jet is assumed to fragment outside the medium. As a result, all formalisms use a medium modified fragmentation function, 
which uses a vacuum fragmentation function as input.  While in the ASW and the HT formalisms, the modification 
is computed in both deconfined and confined phases, due to the assumptions made in the AMY formalism, the modification in 
this formalism occurs only in the deconfined phase. The modification in the confined phase is assumed to be 
small and ignored. While both the HT and the ASW formalisms include contributions from interference with vacuum 
radiation, these are ignored in the AMY scheme. The AMY approach however includes contributions from thermal 
feedback which has so far not been straightforwardly included in the HT and ASW formalisms. The consistent setup of 
the AMY formalism also allows for the most natural extension to include elastic energy loss~\cite{Qin:2007rn}. 
While in the strict interpretation of heavy scattering centers in the ASW (and GLV) formalism, elastic energy loss is
identically zero, the inclusion of elastic loss requires additional assumptions about the medium in the HT approach.
Extensions to the GLV formalism to include mobile scattering centers, and thus, both include elastic energy loss~\cite{Wicks:2005gt} and 
modify the formulation of radiative energy loss~\cite{Djordjevic:2007at} are currently underway. Similar extensions in the HT approach are 
also being carried out~\cite{Wang:2006qr}. However, given the incomplete setup in different formalisms, elastic energy loss was not discussed in these proceedings, 
nor will be included in the realistic comparisons presented in Ref.~\cite{bass}.

While the description of the different formalisms in these proceedings have not included the effect of a dynamical 
medium, realistic calculations of jet modification in heavy-ion collisions do include such effects. The modification 
depends on the path traversed by a given jet. 
This in turn depends on the origin of the jet and the direction of 
travel in the medium. The details related to this problem, as it applies to the different 
formalisms will be presented in the companion paper~\cite{bass}. As a result, comparisons to experimental 
data will also be carried out in this reference as well.

\section{Acknowledgments} 
This work was supported in part by the U.~S.~Department of Energy, under grant numbers DE-FG02-05ER41367 and DE-FG02-01ER41190.
The author thanks S.~A.~Bass, C.~Gale, B.~M\"{u}ller, C.~Nonaka, G.-Y.~Qin, T.~Renk and J.~Ruppert for discussions. 
The author thanks U.~Heinz for a careful reading of the manuscript and for discussions.

\end{document}